\documentclass[preprint,showpacs,preprintnumbers,amsmath,amssymb]{revtex4}
\usepackage[sort&compress]{natbib}
\usepackage{graphicx}
\usepackage{epsfig}

\begin{document}

\title{Effect of time delay on the onset of synchronization of the stochastic Kuramoto model}

\author{Monoj Kumar Sen}
\affiliation{Department of Chemistry,
Visva-Bharati, Santiniketan 731235, India}

\author{Bidhan Chandra Bag{\footnote{Electronic address
:bidhanchandra.bag@visva-bharati.ac.in}}}
\affiliation{Department of Chemistry,
Visva-Bharati, Santiniketan 731235, India}

\author{Karen G. Petrosyan}
\affiliation{Institute of Physics, Academia Sinica, Nankang,
Taipei 11529, Taiwan}

\author{Chin-Kun Hu{\footnote{Electronic address
:huck@phys.sinica.edu.tw}}}
\affiliation{Institute of Physics, Academia Sinica, Nankang,
Taipei 11529, Taiwan} \affiliation{Center for Nonlinear and
Complex Systems and Department of Physics, Chung-Yuan Christian
University, Chungli 320, Taiwan}
\date{\today}

\begin{abstract}
We consider the Kuramoto model of globally coupled phase
oscillators with time-delayed interactions, that is subject to the
Ornstein-Uhlenbeck (Gaussian) colored or the non-Gaussian colored
noise. We investigate numerically the interplay between the
influences of the finite  correlation time of noise $\tau$ and the
time delay $\tau_{d}$ on the onset of the synchronization process.
Both cases for identical and nonidentical oscillators had been
considered. Among the obtained results for identical oscillators
is a large increase of the synchronization threshold as a function
of time delay for the colored non-Gaussian noise compared to the
case of the colored Gaussian noise at low noise correlation time
$\tau$. However, the difference reduces remarkably for large noise
correlation times. For the case of nonidentical oscillators, the
incoherent state may become unstable around the maximum value of
the threshold (as a function of time delay) even at lower coupling
strength values in the presence of colored noise as compared to
the noiseless case. We had studied the dependence of the critical
value of the coupling strength (the threshold of synchronization)
on given parameters of the stochastic Kuramoto model in great
details and presented results for possible cases of colored
Gaussian and non-Gaussian noises.
\end{abstract}


\maketitle

\section{Introduction}

Since the pioneering works on coupled phase oscillators by Winfree
\cite{winfree} and Kuramoto \cite{kuramoto}, synchronization in
nonlinear systems has been systematically studied and attracted
much attention. Recent reviews on the developments can be found in
\cite{strogatz, pikovsky, mikhailov, acebron}. The systems where
the phenomena of synchronization have been observed include
biological clocks \cite{winfree}, chemical oscillators
\cite{kuramoto}, coupled map lattices \cite{kaneko, pecora,gade},
coupled random frequency oscillators \cite{07TangLH},
cardiorespiratory coupled system \cite{06pre-syn}, etc.

The model first introduced by Kuramoto \cite{kuramoto} is one of
the basic models that describes the synchronization process when
initially independent oscillators begin to move coherently. It was
thoroughly studied and successfully applied in several systems
which were modeled by an ensemble of coupled phase oscillators
\cite{acebron}. Another model of interest is the Kuramoto model
with a time-delay \cite{niebur,nakamura,yeung,choi et al}. The
model shows a number of interesting phenomena including, e.g., the
effect of bistability as discovered in \cite{yeung} where the
Kuramoto model with time-delayed interactions was also considered
to be subject to a white noise. Note that both time-delayed
interactions and noise play very important role in nature (see,
e.g., \cite{haken}).

In this paper we consider the Kuramoto model of globally coupled
(all-to-all) phase oscillators with a time delay and subject to
the Ornstein-Uhlenbeck (OU) colored noise \cite{gardiner}, that is
a Gaussian process with a finite correlation time, and a
non-Gaussian colored noise \cite{fuen,06preBag,07preBag,bph}. The
focus in our study is on the interplay between the influence of
noise and the time delay on the synchronization process. Our
previous study on the stochastic Kuramoto model \cite{bph} showed
that the influence of the OU noise qualitatively differs from the
case of white noise as the former allows for the full
synchronization despite the fact that the system is subject to a
noise, see e.g. Fig. 3 in \cite{bph}.

Generally speaking, the time delay introduces de-phasing among the
oscillators. This de-phasing is to interfere with the intrinsic
correlations caused by the finite correlation time of the noise.
As a result, we have found that the de-phasing plays important
role in the dynamics of the system when the intrinsic correlations
are small. We also investigate the effect of the time delay on the
onset of synchronization for different noise strengths for both OU
and non-Gaussian noises. But before doing that we show that the
effect of time delay for the noiseless Kuramoto model is
qualitatively equivalent to the effect of frequency fluctuations
of the phase oscillators. Meanwhile we demonstrate that for the
stochastic Kuramoto model the critical coupling grows nonlinearly
with the increase of the time delay that is due to the additional
de-phasing caused by the time delay. We also investigate how the
transition from the Gaussian to a non-Gaussian noise changes the
dynamical properties of the system.

\section{Stochastic Kuramoto model with time-delayed
interactions}

Let us consider the stochastic Kuramoto model with time-delayed
interactions. This model describes $N$ coupled phase oscillators
with dynamics governed by the following equations
\begin{eqnarray}
\frac{d\theta_i(t)}{dt} = \omega_i + \frac{\epsilon}{N} \sum_{i<j}
^N \sin[\theta_j(t-\tau_d) - \theta_i(t)] + \eta_i (t),
\label{stochastic}
\end{eqnarray}
where $\theta_i$ and $\omega_i$ are, respectively, the phase and
the frequency of the $i$-th oscillators ($i=1,\dots,N$),
$\epsilon$ is the coupling constant, and $\tau_d$ is the time
delay. The independent noise processes $\eta_i (t)$ are governed
by
\begin{eqnarray}
\frac{d\eta_i (t)}{dt} = - \frac{1}{\tau}\frac{d}{d \eta_i} U_p(\eta_i)
+ \frac{\sqrt{D}}{\tau} \xi_i (t). \label{noise}
\end{eqnarray}
The potential function is
\begin{eqnarray}
U_p(\eta_i)=\frac{D}{\tau( p-1)} \ln[1+\alpha (p-1)\eta_i^2/2]
\nonumber
\end{eqnarray}
with $\alpha = \tau / D$. $\xi_i(t)$ is the Gaussian white noise
process defined via $$\langle\xi_i(t)\xi_j(t')\rangle =
2\delta_{ij}\delta(t-t')$$ and $\langle\xi_i(t)\rangle = 0$. $D$
and $\tau$ measure the intensity and the correlation time of the
noise process. Shortly we will discuss more about them. In the
noise term in Eq. (1) it is apparent that in the present problem
we assume the homogenous diffusion of phase oscillators.

The form of the noise $\eta_i$ allows us to control the deviation
from the Gaussian behavior by changing a single parameter $p$. For
$p=1$,  Eq.(\ref{noise}) becomes
\begin{equation}
\frac{d\eta_i}{dt} = - \frac{\eta_i}{\tau} + \frac{\sqrt{D}}{\tau}
\xi_i (t)
\end{equation}
which is a well-known time evolution equation for the OU noise
process \cite{gardiner} for which the auto correlation function is given by
\begin{equation}
<\eta_i(t)\eta_i(t')> = \frac{D}{\tau}\exp(-\frac{|t-t'|}{\tau}) \; \;.
\label{gvar}
\end{equation}

\noindent Thus $D$ and $\tau$ are the noise strength and
correlation time of the OU noise. The factor $[1+\alpha
(p-1)\eta_i^2/2]$ in  the first term of Eq.(\ref{noise}) leads to
producing  colored non Gaussian noise whose effective noise
strength and correlation time are different from $D$ and $\tau$.
However, to make the present paper self-consistent we would like
to mention here salient features of the non-Gaussian noise only
\cite{fuen}.

The stationary probability distribution of the noise process is given by
\begin{equation}
P(\eta_i)=\frac{1}{Z_{ip}}\left[1+\alpha (p-1)
\frac{\eta_i^2}{2}\right]^{\frac{-1}{p-1}},
\label{df}
\end{equation}
where $Z_{ip}$ is the normalization constant which equals to
\begin{eqnarray}
Z_{ip}  =  \int_{-\infty}^{\infty}d\eta_i \left[1+\alpha (p-1)
\frac{\eta_i^2}{2}\right]^{\frac{-1}{p-1}}  \nonumber\\
 =  \sqrt{\frac{\pi}{\alpha
 (p-1)}}\frac{\Gamma_1(1/(p-1)-1/2)}{\Gamma_1(1/(p-1))} \; \;
\;,\label{nc}
\end{eqnarray}
with $\Gamma_1$ being the Gamma function. This distribution
can be normalized only for $p<3$. Since $P(\eta_i)$ is an even
function of $\eta_i$, the first moment, $\langle \eta_i \rangle$, is
always 0, and the second moment (variance) is given by

\begin{equation}
\langle \eta_{ip}^2 \rangle  = \frac{2 D}{\tau (5-3 p)} \; \;
\;,\label{var}
\end{equation}
which is finite only for $p<5/3$. Furthermore, for $p<1$, the
distribution has a cut-off and it is only defined for $|\eta_i| <
\eta_{ic} \equiv \sqrt{\frac{2D}{\tau (1-p)}}$.

However, it is difficult to determine the auto correlation function (ACF) of
the non Gaussian noise $\eta_i$ exactly. To have an idea about this we have
 calculated it numerically and presented the result in Fig. 1.
The two time correlation function for the non Gaussian noise
(solid curve) is fitted well by bi-exponentially decaying function
$<\eta_i(t)\eta_i(0)>= a e^{-t/37}+ b e^{-t}$ having two
correlation times 37 and 1, respectively, for $\tau=1.0$, where
$a$ and $b$ are constants related to the noise intensity. In the
same figure we have plotted numerically calculated ACF for
Gaussian colored noise. It exactly mimics the function given in
Eq.(\ref{gvar}). Figure 1 clearly shows that the effective noise
strength and correlation time for non Gaussian noise are greater
compared to those of colored Gaussian one. The correlation time of
non-Gaussian noise $\tau_p$ at the stationary regime of the
process $\eta_i(t)$ diverges near $p=5/3$ and it can be
approximated \cite{fuen} over the range  $1 \le p < 5/3$ as

\begin{equation}\label{ct}
\tau_p \simeq 2\tau/(5-3p) \; \;.
\end{equation}

This equation is qualitatively consistent with Fig.1. However, for
this approximate correlation time, Eq.(\ref{var}) becomes

\begin{eqnarray}
\langle \eta_{ip}^2 \rangle \simeq  \frac{4 D}{\tau_p (5-3 p)^2}
\label{var1} \; \; \;.
\end{eqnarray}

Equations (\ref{var}-\ref{var1}) imply that when $p\rightarrow 1$,
we recover the limit that $\eta_i$ is a Gaussian colored noise
since at this limit Eqs.(\ref{var},\ref{var1}) correspond to the
variance of Colored Gaussian noise as given in Eq.(\ref{gvar}) and
$\tau_p$ in Eq.(8) becomes $\tau$ which is the correlation time of
the Gaussian noise. Another check in this context can be obtain in
the following way. In this limit the term in the square bracket of
Eq.(\ref{df}) can be written as

\begin{equation}
1+\alpha
(p-1){\eta_i^2}/{2}=\exp(\alpha(p-1){\eta_i^2}/{2}) \; \;.
\end{equation}

\noindent
Then Eq.(\ref{df}) becomes

\begin{equation}
P(\eta)=\frac{1}{Z_1}\exp(-\alpha \eta_i^2/2) \; \; \;, \label{eq9}
\end{equation}
with $Z_1=\sqrt{\pi/\alpha}$, which is a Gaussian distribution
function. However, Eq.(\ref{var}) shows that for a given external
noise strength $D$ and the noise correlation time $\tau$, the
variance of the non-Gaussian noise is higher than that of the
Gaussian noise for $p>1$, {\it i.e.} $\langle \eta_{ip}^2 \rangle
> \langle \eta_{i1}^2 \rangle$. Similarly, Eq.(\ref{ct}) implies
that $\tau_p > \tau$ for $p>1$. These are consistent with the
message of Fig. 1. Before leaving this issue we like to emphasize
that in the present study we have considered continuous
distribution of the non-Gaussian noise which is more relevant to
the natural systems rather than two-state or discrete
distributions as mostly used in the literature to study the noise
driven dynamical systems \cite{doer}.

The quantity of interest in the present study is
\begin{eqnarray}
Z = \Gamma e^{i\Theta} = \frac{1}{N} \sum_{i=1} ^{N} e^{i\theta_i}
\label{orderparameter}
\end{eqnarray}
which is the order parameter that measures the extent of
synchronization in the system of $N$ phase oscillators. Its
absolute value $\Gamma$ determines the degree of synchronization.
It can be seen that in case of all the oscillators having the same
phase the quantity equals to one ($\Gamma = 1$) that corresponds to
the full synchronization. The degree of synchronization is equal
to zero ($\Gamma = 0$) when all the oscillators are independent and
have different phases. $\Theta$ defines an average phase of the
oscillators. For the initial distribution of frequencies we choose the
Lorentzian distribution
\begin{eqnarray}
g(w) = \frac{1}{\pi}\frac{\lambda}{(\omega - \omega_0)^2 +
\lambda^2}. \label{lorentz}
\end{eqnarray}

\section{Results and discussion}


We have computed the order parameter $\Gamma$ as well as the
critical coupling strength numerically since it is very difficult
to study the problem analytically due to the colored noise and
nonlinearity in terms of $\eta_i$ in its time evolution equation.
Using Heun's method (a stochastic version of the Euler method
which reduces to the second order Runge-Kutta method in the
absence of noise) we have solved the Eqs. (\ref{stochastic}) and
(\ref{noise}) simultaneously \cite{rt}. Based on the above method,
we have studied time evolution of $N=5000$ coupled phase
oscillators. In Ref. \cite{bph}, we have showed that such a $N$
value is large enough so that the calculated quantities can
represent those for systems in the thermodynamic limit.

For a given frequency distribution of the phase oscillators and
environment of specific characteristic there exists a threshold
value of the coupling strength $(\epsilon_c)$ above which the
coherent (synchronized) state is the stable state. If the
$\epsilon$ is smaller than $\epsilon_c$, then the incoherent state
is the stable one. $\epsilon_c$ is called critical coupling
strength. To calculate the value of $\epsilon_c$, we first obtain
the dependence of ensemble average of $\Gamma$ ($<\Gamma>$) on
$\epsilon$. From this we determine numerically the derivative of
$<\Gamma>$ with respect to $\epsilon$ at stationary state. The
$\epsilon$ corresponding to the maximum value of the derivative is
identified as the critical coupling strength. Our numerical study
satisfies all the known limiting results, i.e., (i) $\epsilon_c=
2\lambda$ in the absence of noise ($D=0$) and $\tau_d=0$  (ii)
$\epsilon_c=2 (\lambda+ D)$ for white noise and $\tau_d=0$
(results are not shown here). We would also like to mention here
that we chose stationary time to be a linear function of time
delay ($\tau_d$) with the proportionality constant 100 and
integration step length $h=0.01$.

We start our numerical study to determine how stability of both
coherent and incoherent states depends on the time delay.
Stability of these states of white-noise driven coupled
oscillators in the Kuramoto model with a time delay has been
studied by Young and Strogatz \cite{yeung}. They have considered
two cases. In the first case, all the oscillators are identical.
In the second case, they have chosen Lorentzian distribution,
Eq.(\ref{lorentz}), of frequency of the coupled oscillators. In
the present paper, we extend this study to colored-noise driven
coupled oscillators. The noise may be OU Gaussian or non Gaussian
in characteristics. However, the critical coupling strength as
mentioned above indicates that up to that value of coupling
strength, the incoherent state can survive at long time. This
range corresponds the stability zone of the incoherent state of
the coupled phase oscillators. Thus critical coupling strength is
a measure to imply how stability of incoherent or coherent states
of the coupled oscillators depends on time delay and other
parameters of the system. To demonstrate the dependence of the
critical coupling strength $\epsilon_c$ on the time delay $\tau_d$
for various noise properties, we  have reproduced Fig.2 of Ref.
\cite{yeung} for identical oscillators in the absence of noise
(i.e. $D=0$) and presented it by the solid curve in Fig.2. In the
presence of noise the phase oscillators become non identical and
the time delay is effective in the dynamics even at its low value
compared to the case where noise is absent. Since the variance of
the non-Gaussian noise is higher compared to the Gaussian one, the
shift of the damped oscillating curve towards the larger critical
coupling strength from the noiseless case, is very big for former
than latter at low noise correlation time, $\tau \approx 0.1$.
Their difference reduces remarkably at large noise correlation
time, $\tau \approx 2$. The reason may be the following. Effective
noise correlation time of the colored non-Gaussian noise is much
higher than the Gaussian one at large $\tau$. Increase of noise
correlation time leads to develop better phase relationship among
the oscillators reducing the phase diffusion. Thus at large $\tau$
the difference in critical coupling strength is small for a given
time delay. It is apparent in Fig.2 that the interplay of noise
correlation time and the time delay plays some constructive role
to have a synchronized state particularly at large $\tau$ and
around the first maximum of the damped oscillation curve. It also
shifts a little bit the position of the maximum towards the left
and reduces the oscillation amplitude at large time delay values.
Thus colored noise plays a role beyond the induction of diffusion
behavior.

 Before going to the next case we would like to demonstrate the
variation of critical coupling strength as a function of frequency
of the coupled identical oscillators in the presence of time
delay. In Fig.3 we have presented this. It exhibits that the
variation is periodic as a result of interplay of the time delay
and the frequency of the oscillator. However, again it shows how
noise correlation can reduce the stability of incoherent state.
Because of higher effective noise strength of the non-Gaussian
noise, the critical coupling strength for the case $p=1.5$ and
$\tau=0.5$ is comparable to white Gaussian noise.

 For the second case of non-identical phase oscillators,
 we have done calculations similar to those for Fig. 2
for identical oscillators, and the results are presented in Fig.
4. Again, the solid curve in Fig.4 is a reproduction of Fig. 4 of
Ref.\cite{yeung}. All the features of Fig. 2 in the presence of
colored noise has appeared in Fig. 4. Here one important point to
be noted is that around the maximum, the incoherent state is
unstable even at lower coupling strength in the presence of
colored noise compared to the case without noise. Thus colored
noise can effect the nonlinear coupling among the phase
oscillators. It is quite similar to the modification of the
dynamics by the colored noise in the presence of nonlinear
potential \cite{han1,mb}.

 In the next step we demonstrate how the critical coupling
strength varies as a function of noise parameters. First, we
consider the dependence of $\epsilon_c$ on the noise strength (D).
To identify the signature of the time delay in this context we
have determined a ratio of critical coupling strengths
$\epsilon_c(\tau_d=0)/\epsilon_c(\tau_{dm})$ for a given noise
intensity, where $\tau_{dm}$ is the location of the first maximum
in the $\tau_d$ axis. For identical oscillators we have chosen
$\tau_{dm}=2$ from Fig. 2 and for nonidentical oscillators we have
considered $\tau_{dm}=1$ from Fig. 4.
It is not difficult to
anticipate that this choice will incorporate the maximum effect of
time delay. However, the ratio had been calculated for different
values of the noise strength for the white noise and is presented
in Fig.5. At first it rapidly increases both for identical (dotted
curve) and nonidentical (solid curve) oscillators then the growth
rate slows down. The initial growth rate is higher for identical
oscillators compared to nonidentical oscillators because at low
noise strength $\epsilon_c(\tau_d=0)$ is close to zero for the
former and it grows at a faster rate. If the noise strength is
appreciably large then the ratio is higher for identical
oscillators than for the nonidentical case even if $\tau_{dm}$ is
greater for the former than the latter. Thus time delay is more
effective in the dynamics for nonidentical oscillators compared to
the case of identical oscillators. Similar features are also
observed for the colored noise and are demonstrated in insets (a)
and (b) for the Gaussian and non-Gaussian noises. At large noise
intensities the ratio of critical coupling strengths is smaller
for the colored Gaussian noise compared to other noises for the
case of nonidentical oscillators. Thus on switching from the white
Gaussian to colored Gaussian or non-Gaussian characteristics of
noise the time delay begins to play an important role in the
dynamics.

 The variation of the ratio of critical coupling strengths with
the non-Gaussian parameter $p$ is demonstrated in Fig.6. It shows
that the ratio increases as a function of $p$ for both cases of
identical and nonidentical oscillators. The rate of growth is
higher for the former than the latter case. The ratio is lower for
the identical oscillators compared to the case of nonidentical
oscillators at low $p$. However at large $p$ the situation turns
around, it becomes inverse. These aspects can be understood if one
keeps in mind that the effective noise strength increases as $p$
grows. Therefore Fig.5 is essentially similar to Fig.6 and it can
be explained in the same way as we have explained the results
presented in Fig.6.

 Finally, in Fig.7 we have presented the variation of the ratio
of critical coupling strength as a function of noise correlation
time. It shows that the ratio decreases with increase of $\tau$.
Since phase diffusion reduces as the noise becomes more colored,
the critical coupling strength decreases as $\tau$ grows both in
the presence and the absence of time delay. The ratio decreases
because in the absence of time delay the colored noise is more
effective in the dynamics. Similarly, the rate of decrease is a
little bit higher for identical oscillators than for nonidentical
ones. These kind of features are also observed for the colored
non-Gaussian noise and are presented in the inset of the Fig. 7.
In both cases the ratio is higher for the identical oscillators
compared to the nonidentical oscillators for a given noise
correlation time. Thus it again supports the statement that the
time delay is more effective in the dynamics for the latter than
for the former.

\section{Conclusion}

We have considered the stochastic Kuramoto model including
time-delayed interaction with a delay time $\tau_d$ subject to
both OU Gaussian or non-Gaussian colored noise with a correlated
time $\tau$. The main focus of our study was on the interplay
between the effects of finite correlation time $\tau$ and the time
delay $\tau_d$ and their influence on the onset of the
synchronization. Our results can be summarized as follows.

(i) The shift of the damped oscillating curve (for the critical
coupling strength as a function of time delay $\tau_d$) towards
the larger critical coupling strength from the noiseless case is
very large for the non-Gaussian noise than for the Gaussian one at
low noise correlation time, $\tau \approx 0.1$. Their difference
reduces remarkably at large noise correlation times, $\tau \approx
2$ (Fig. 2).

(iii) Due to the present intrinsic correlations, the colored noise
plays a role beyond the induction of diffusive behavior.

(iii) The incoherent (unsynchronized) state may be unstable around
the maximum even at lower coupling strength values in the presence
of colored noise compared to the noiseless case (Fig. 4).

(iv) The ratio of critical coupling strengths
$\epsilon_c(\tau_d=0)/\epsilon_c(\tau_{dm})$ first rapidly
increases as a function of noise intensity $D$ for both the
identical and nonidentical oscillators and then slows down. The
initial growth rate is higher for identical oscillators compared
to the case of nonidentical oscillators. If the noise strength is
appreciably large then the ratio is higher for identical
oscillators than for the nonidentical case even if $\tau_{dm}$ is
greater for the former than the latter case. At large values of
noise intensity the ratio of critical coupling strength is smaller
for the colored Gaussian noise than for the other noises for the
case of nonidentical oscillators (Fig. 5).

(vi) The ratio of critical coupling strengths increases as a
function of non-Gaussian parameter $p$ for both cases of identical
and nonidentical oscillators. The rate of growth is higher for the
former case than the latter one. The ratio is lower for identical
oscillators compared to nonidentical oscillators at low $p \ge 1$.
However at large values of $p$ it becomes inverse (Fig. 6).

(v) The above ratio decreases with increase of $\tau$. The rate of
decrease is higher for the case of identical oscillators compared
to the one for nonidentical oscillators (Fig. 7).

 We anticipate that investigation of influence of time-delayed
interactions and (generally non-Gaussian) noise in complex systems
would lead to important insights into their stochastic dynamics,
e.g., for the process of intercellular synchronization in biology
\cite{zhou} (see also systems biology models mentioned as possible
applications in \cite{bph}). The time-delayed interaction also
plays an important role in a molecular model of biological
evolution \cite{10pre06evo}.

This work was supported by the National Science Council in Taiwan
under Grant Nos. NSC 96-2911-M 001-003-MY3 \& NSC
98-2811-M-001-066, and National Center for Theoretical Sciences in
Taiwan.

\newpage

\begin{figure}
\includegraphics[width=1.00\linewidth,angle=0]{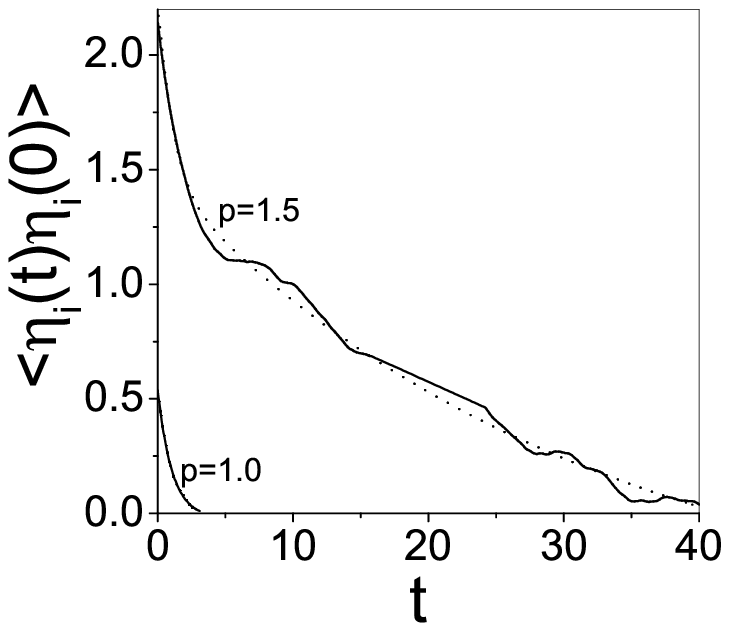} \hfill
\caption{Two-time correlation function {\it vs.} $t$ of both
Gaussian ($p=1$) and non-Gaussian ($p=1.5$) colored noises for
parameters $\tau=1.0$ and $D=0.5$.}\label{fig1}
\end{figure}

\begin{figure}
\includegraphics[width=1.00\linewidth,angle=0]{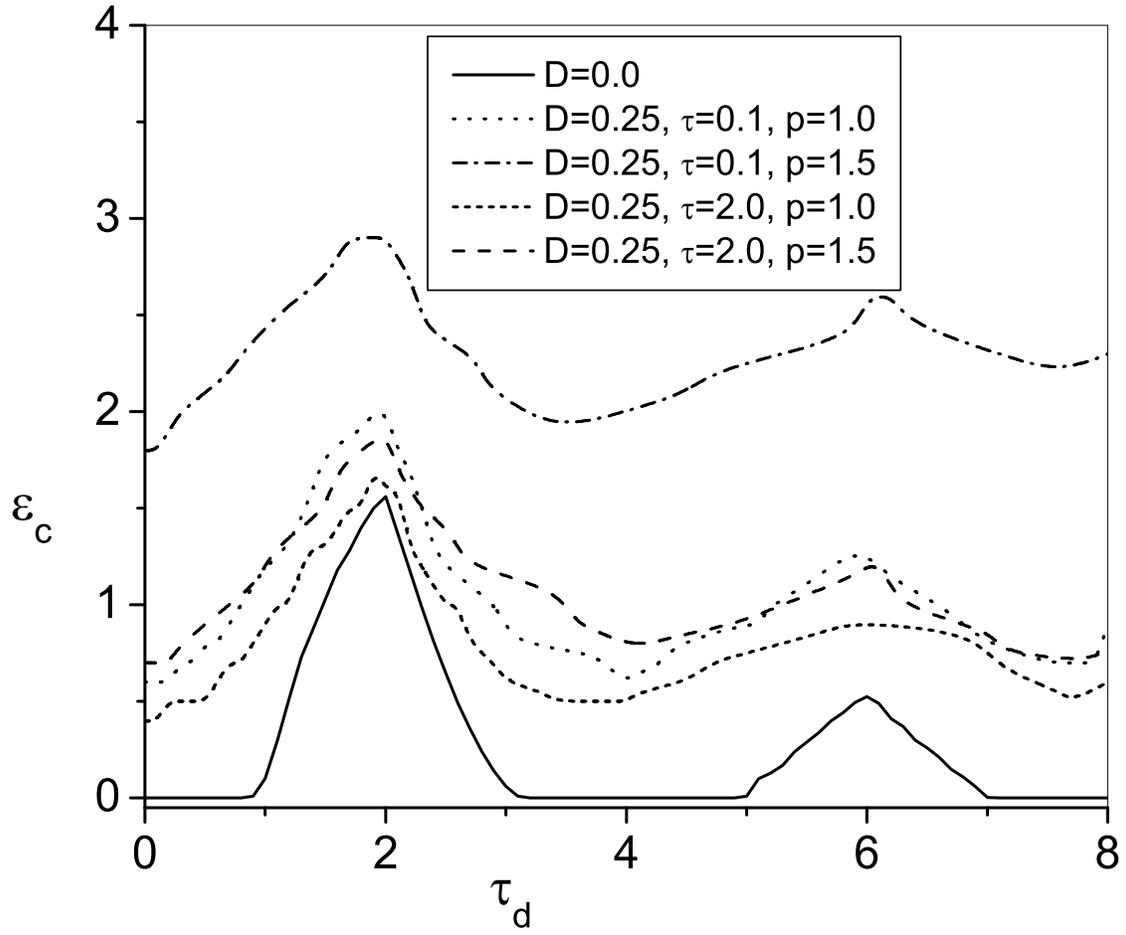} \hfill
\caption{Plot of the critical value of coupling strength
$\epsilon_c$ vs the time delay $\tau_d$ for the identical
oscillators for $\omega_0=\pi/2$. } \label{fig2}
\end{figure}

\begin{figure}
\includegraphics[width=1.00\linewidth,angle=0]{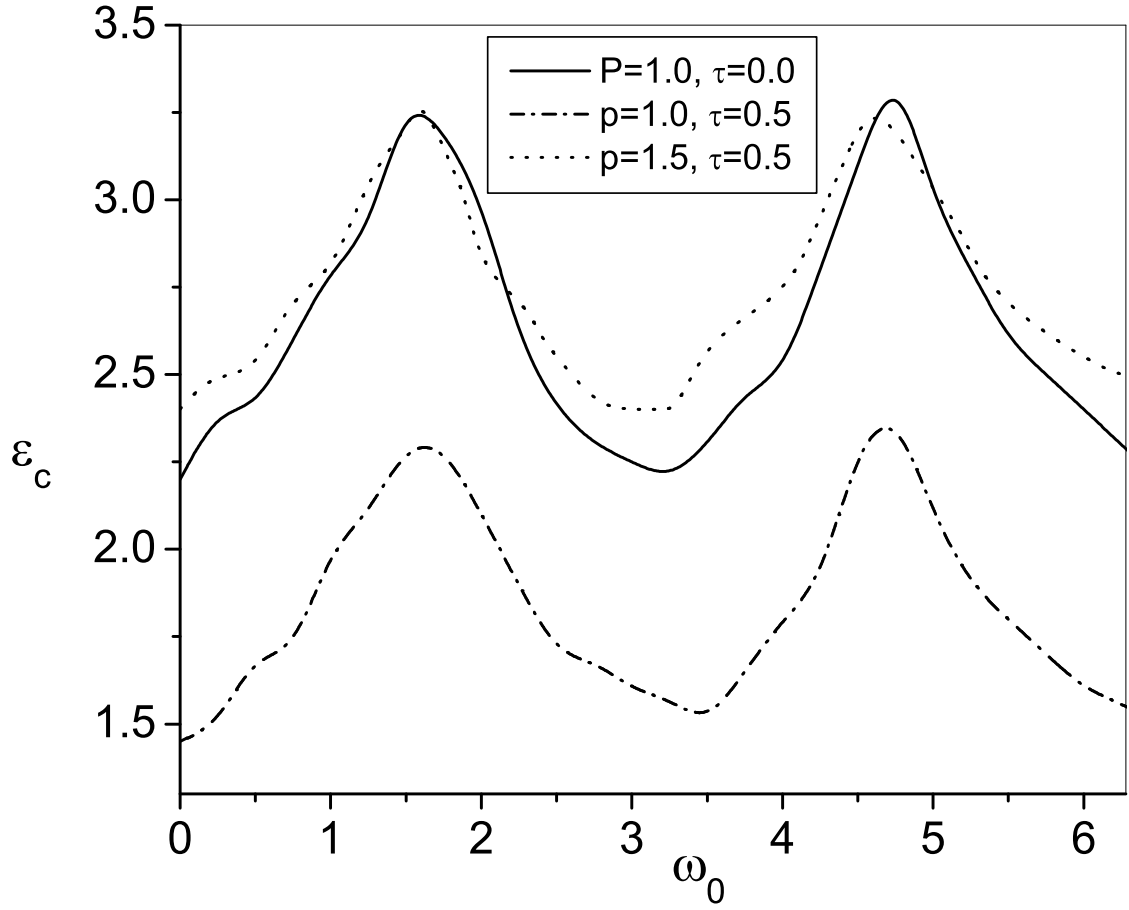} \hfill
\caption{Plot of the critical value of coupling strengths
$(\epsilon_c)$ vs the frequency of identical oscillators
$\omega_0$ for $D=1.0$ and $\tau_d=2.0$.} \label{fig3}
\end{figure}

\begin{figure}
\includegraphics[width=1.00\linewidth,angle=0]{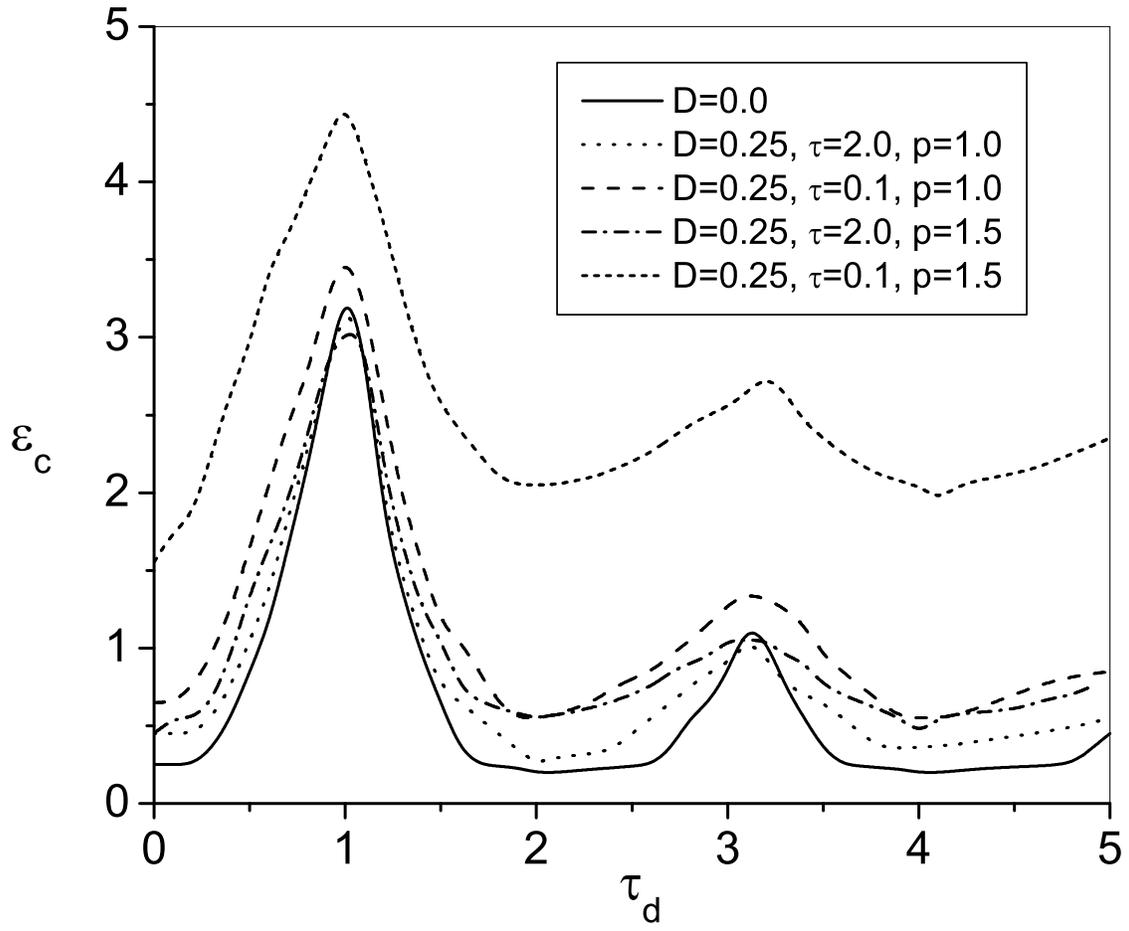} \hfill
\caption{Plot of the critical value of coupling strength
$\epsilon_c$ vs the time delay $\tau_d$ for the nonidentical
oscillators for the parameter set $\omega_0=3.0$ and
$\lambda=0.1$. } \label{fig4}
\end{figure}

\begin{figure}
\includegraphics[width=1.00\linewidth,angle=0]{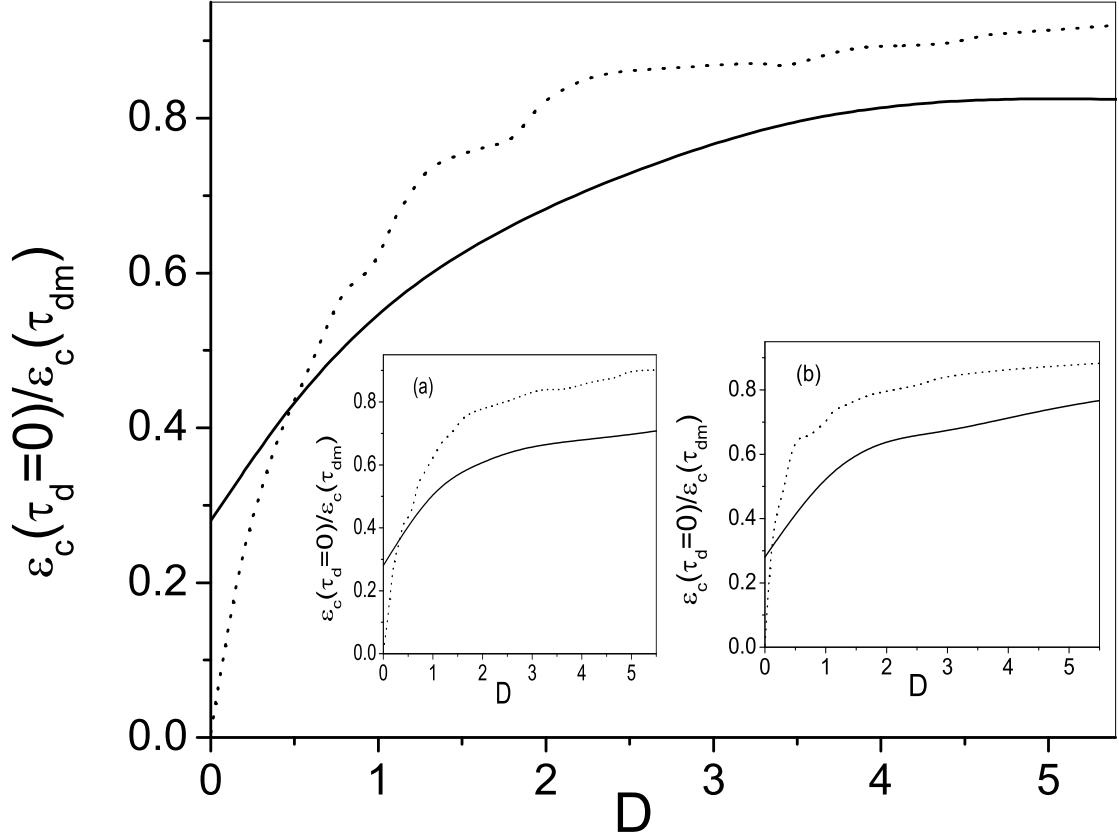} \hfill
\caption{Plot of the ratio of the critical value of coupling
strengths $\epsilon_c(\tau_d=0)/\epsilon_c(\tau_{dm})$ vs the
noise intensity $D$ for the parameter set $p=1.0$ and $\tau=0.0$.
For identical oscillators $\omega_0=\pi/2$ and that of
nonidentical oscillators, $\omega_0=3.0$, $\lambda=0.5$. Solid and
dotted curves correspond to nonidentical and identical
oscillators, respectively. The insets contain results for
different parameter sets: (a) $p=1.0$ and $\tau=0.5$, (b) $p=1.5$
and $\tau=0.5$} \label{fig5}
\end{figure}

\begin{figure}
\includegraphics[width=1.00\linewidth,angle=0]{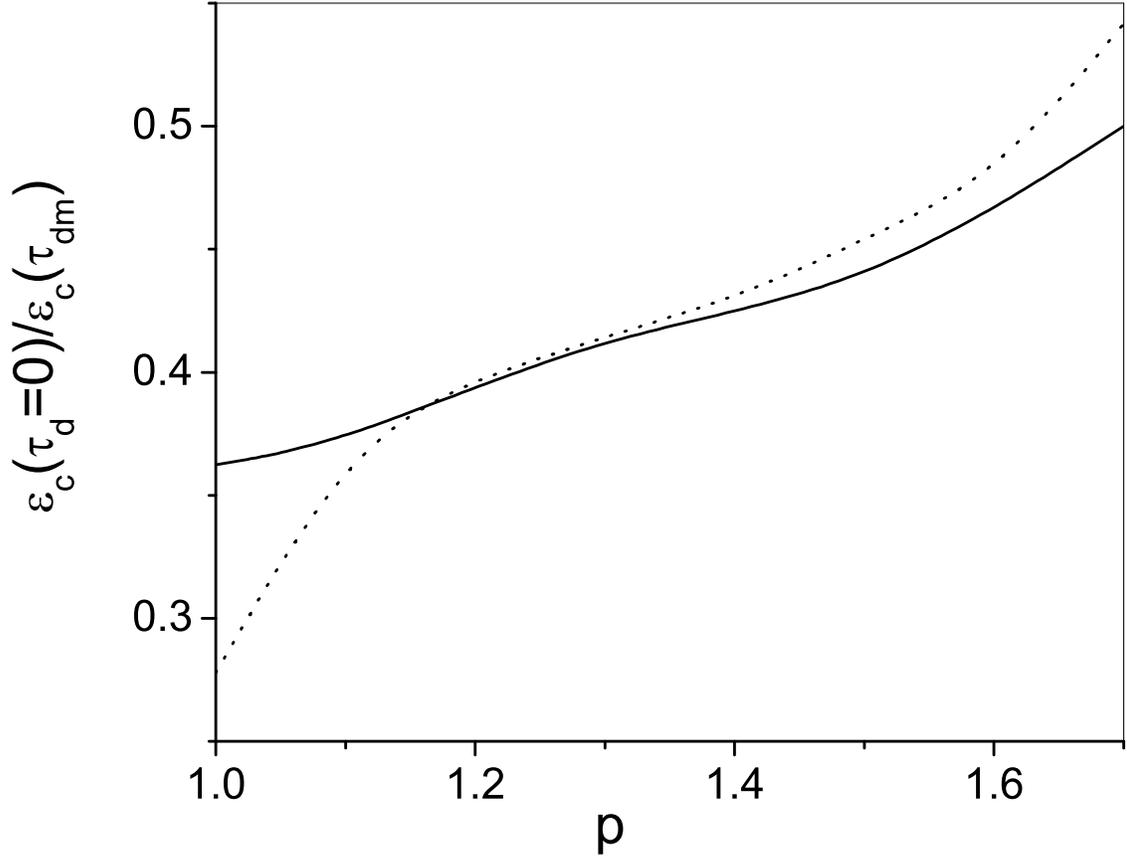} \hfill
\caption{Plot of the ratio of the critical value of coupling
strengths $\epsilon_c(\tau_d=0)/\epsilon_c(\tau_{dm})$ vs the
non-Gaussian parameter $p$ for the parameter set $\tau=0.5$ and
$D=0.25$. For identical oscillators $\omega_0=\pi/2$ and that of
nonidentical oscillators $\omega_0=3.0$ and $\lambda=0.5$. Solid
and dotted curves correspond to nonidentical and identical
oscillators, respectively.} \label{fig6}
\end{figure}

\begin{figure}
\includegraphics[width=1.00\linewidth,angle=0]{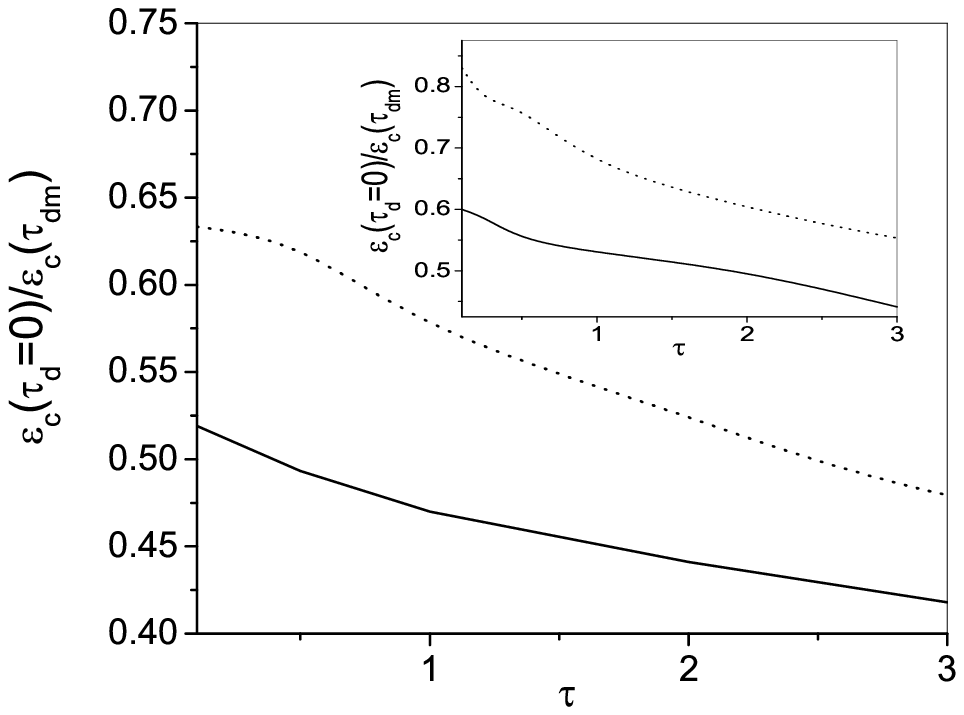} \hfill
\caption{Plot of the ratio of the critical value of coupling
strengths $(\epsilon_c(\tau_d=0)/\epsilon_c(\tau_{dm}))$ vs the
noise correlation time $\tau$ for the parameter set $p=1.0$ and
$D=1.0$. For identical oscillators $\omega_0=\pi/2$ and that of
nonidentical oscillators $\omega_0=3.0$ and $\lambda=0.5$. Solid
and dotted curves correspond to nonidentical and identical
oscillators, respectively. In the inset, $p=1.5$.} \label{fig7}
\end{figure}


\begin{thebibliography}{99}

\bibitem{winfree} Winfree A T 1967 J. Theor. Biol. {\bf 16} 15

Winfree A T 1980 {\it The Geometry of Biological Time} (Springer,
New York)

\bibitem{kuramoto}Kuramoto Y 1984 Prog. Theor. Phys. Suppl. {\bf 79} 223

Kuramoto Y 1984{\it Chemical Oscillations, Waves, and Turbulence}
(Springer-Verlag, New York)

Kuramoto Y and Nishikawa I 1987 J. Stat. Phys. {\bf 49} 569


\bibitem{strogatz} Strogatz S H 2000 Physica (Amsterdam) {\bf 143D} 1


\bibitem{pikovsky} Pikovsky A, Rosenblum M, and Kurths J 2001 {\it Synchronization - A
Universal Concept in Nonlinear Sciences} (Cambridge University
Press, Cambridge)

\bibitem{mikhailov}Manrubia S C, Mikhailov A S and
Zanette D H 2004 {\it Emergence of Dynamical Order:
Synchronization Phenomena in Complex Systems} (World Scientific,
Singapore)

\bibitem{acebron}Acebron J J, Bonilla L L, Perez-Vicente C J, Ritort F and
Spigler R 2005 Rev. Mod. Phys. {\bf 77} 137

\bibitem{kaneko}Kaneko K 1990 Phys. Rev. Lett. {\bf 65} 1391

\bibitem{pecora}Pecora L M and Caroll T L 1998 Phys. Rev. Lett. {\bf 80} 2109

\bibitem{gade}Gade P M and Hu C-K 1999 Phys. Rev. E. {\bf 60} 4966

Gade P M and Hu C-K 2000 Phys. Rev. E. {\bf 62} 6409

Gade P M and Hu C-K 2006  Phys. Rev. E. {\bf 73} 036212

Jalan S, Amritkar R E and  Hu C-K 2005  Phys. Rev. E. {\bf 72}
016211

Amritkar R E, Jalan S and C.-K. Hu C-K 2005 Phys. Rev. E {\bf 72}
016212



Hung Y-C, Huang Y-T, Ho M-C, and Hu C-K 2008 Phys. Rev. E {\bf 77}
016202


\bibitem{07TangLH}
 Hong H, Choi M Y and  Kim B J 2002 Phys. Rev. E {\bf 65} 026139

Hong H, Chate H, Park H and Tang LH 2007 Phys.
Rev. Lett. {\bf 99} 184101

Hong H, Park H and Tang LH 2007 Phys. Rev. E {\bf 76}  066104

\bibitem{06pre-syn} Wu M-C and Hu C-K 2006
Phys. Rev. E {\bf 73} 051917

\bibitem{niebur}Niebur E, Schuster H G, and Kammen D M 1991 Phys. Rev. Lett.
{\bf 67} 2753

\bibitem{nakamura}Nakamura Y, Tominaga F and Munakata T 1994 Phys. Rev. E {\bf 49}
4849

\bibitem{yeung}Yeung M K S and Strogatz S H 1999 Phys. Rev. Lett. {\bf 82} 648

\bibitem{choi et al} Choi M Y, Kim H J, Kim D and Hong H 2000
Phys. Rev. E {\bf 61} 371

\bibitem{haken} Haken H 2002 {\it Brain Dynamics: Synchronization and
Activity Patterns in Pulse-Coupled Neural Nets with Delays and
Noise} (Springer-Verlag, Berlin)

\bibitem{gardiner}Gardiner C W 2004 {\it Handbook of Stochastic
Methods}, 3rd ed. (Springer-Verlag, Berlin)

\bibitem{fuen}See Fuentes M A, Wio H S and Toral R 2002 Physica A {\bf 303}
91  for details on the non-Gaussian noise

\bibitem{06preBag}Bag B C and Hu C-K 2006 Phys. Rev. E {\bf 73} 061107

\bibitem{07preBag}Bag B C and Hu C-K 2007 Phys. Rev. E {\bf 75} 042101

\bibitem{bph}Bag B C, Petrosyan K G and Hu C-K 2007 Phys. Rev. E {\bf 76}
056210

\bibitem{doer} H\"anggi P, Talkner  P and Borkovec M 1990 Rev. Mod. Phys. {\bf 62} 251

Doering C R and J. C. Gadoua J C 1992 Phys. Rev. Lett {\bf 69},
2318

Reimann P 2002 Phys. Rep. {\bf 361} 265

\bibitem{rt}
Toral R 1995 in \textit{Computational Physics, Lecture Notes in
Physics}, Vol.448 (Springer-Verlag, Berlin) Eds. Garrido  P and
Marro J


\bibitem{han1} H\"anggi P 1994 Chem. Phys. {\bf 180} 157

\bibitem{mb} Sen M K and Bag B C 2009 Euro. Phys. J. B {\bf 68} 253

\bibitem{zhou}Zhou T, Chen L and Aihara K 2005 Phys. Rev. Lett. {\bf 95} 178103

\bibitem{10pre06evo}
Saakian D B, Martirosyan A S and Hu C-K 2010 Phys. Rev. E {\bf 81}
061913

\end{thebibliography}
\end{document}